\documentclass[twocolumn,aps,prd,showpacs,superscriptaddress,floats,floatfix,showkeys,notitlepage]{revtex4-1}

\usepackage{enumitem}
\usepackage{times}
\usepackage{dcolumn}
\usepackage{mathrsfs}
\usepackage{hyperref}
\usepackage{amsmath}
\usepackage{mathtools}
\usepackage{bbm}
\usepackage{bm}
\usepackage{amsfonts}
\usepackage{amssymb}
\usepackage{mathrsfs}
\usepackage{wasysym}
\usepackage{graphicx}
\usepackage[]{subfigure}
\usepackage{filecontents}
\usepackage[dvipsnames]{xcolor}
\usepackage{bbold}
\usepackage{lipsum}
\usepackage{graphicx}
\usepackage{subfigure}
\usepackage{palatino}
\usepackage{changes}
\usepackage{hyperref}
\hypersetup{colorlinks=true,linkcolor=blue,urlcolor=blue,citecolor=blue}
\usepackage[toc,page]{appendix}
\usepackage[normalem]{ulem}
\usepackage{adjustbox}
\usepackage{latexsym}
\usepackage{amsmath}
\usepackage{amssymb}
\usepackage{amsfonts}
\usepackage{times}
\usepackage{dcolumn}
\usepackage{bm}
\usepackage{tikz}
\usepackage{bigints}
\usepackage{array,tabularx,multirow}
\usepackage[dvipsnames]{xcolor}
\usepackage[tracking=true]{microtype}
\SetTracking{}{500}
\SetTracking{encoding={*}, shape=sc}{40}
\UseRawInputEncoding 
\allowdisplaybreaks

\usepackage{braket}
\usepackage{tensor}
\usepackage{slashed}
\usepackage{booktabs} 
\usepackage{float}

\begin{document} \sloppy

\title{Observational signatures of Rotating compact objects in Plasma space-time}

\author{Saurabh Kumar}
\email{ksaurabhkumar.712@gmail.com}
\affiliation{Department of Physics, Indian Institute of Technology, Guwahati 781039, India}

\author{Akhil Uniyal}
\email{akhil\_uniyal@sjtu.edu.cn}
\affiliation{Tsung-Dao Lee Institute, Shanghai Jiao Tong University, Shengrong Road 520, Shanghai, 201210, People’s Republic of China}

\author{Sayan Chakrabarti}
\email{sayan.chakrabarti@iitg.ac.in}
\affiliation{Department of Physics, Indian Institute of Technology, Guwahati 781039, India}

\date{\today}
\begin{abstract}
We have investigated the characteristics of shadows cast by the Kerr black hole in the presence of plasma and compared them to those of a rotating wormhole in a uniform plasma space-time for an observer at infinity. Interestingly, for the same uniform plasma density, the apparent shadow size of the rotating wormhole is always greater than that of the Kerr black hole. To further distinguish the two compact objects we studied the deflection angle and did a comparative study in the presence of the uniform and non-uniform plasma profiles. The goal of this whole exercise is to deepen our understanding of the observational phenomena of these astrophysical objects. The analysis reveals the importance of specific plasma distribution profiles, the impact of plasma on the shadow diameter, and the behavior of deflection angles in different plasma scenarios. We have calculated constraints on the plasma parameters by considering observational data and employing analytical formulations. Our work therefore provides valuable insights into the behavior of light rays near compact objects in plasma space-time.
\end{abstract}

\maketitle

\date{\today}


\section{Introduction}
Black holes (BH) and wormholes (WH) are geometrically two different astrophysical objects, however observationally the direct evidence of their existence is still not full-proof to say the least. Although both BH and WH are solutions to Einstein's equations, they are distinctly different. Interestingly, recent shadow images of M87* \cite{EventHorizonTelescope:2019dse} and SgrA* \cite{EventHorizonTelescope:2022xnr} give us hope for the existence of such objects. The wormhole remains mysterious so far, with the only similarity that exists between them being that they are both solutions of Einstein's Field Equations. Wormholes are among the most exciting objects predicted by Einstein's general theory of relativity, which topologically have a tunnel-like structure connecting two different regions of space-time. The first description of such a geometry appeared as early as 1916 \cite{Flamm_1916} in a study of the spatial part of the Schwarzschild metric. The idea was later refined by Einstein and Rosen, which is now known as the Einstein-Rosen bridge \cite{Einstein1935}. However, it was more of a mathematical construction, since it did not deal with traversable WHs. Ellis \cite{Ellis:1973yv} first came up with the traversable general relativistic solution of the WH, and later on, Morris and Thorne \cite{Morris:1988cz} studied the properties of such WHs in detail. Although WHs are hypothetical and topologically non-trivial structures in space-time, it is true that, from a theoretical standpoint, the possibility of their existence is difficult to answer. However, they should not be completely ruled out as they can very well play the role of a black hole mimickers or could be thought of as one of the many different exotic compact objects in the universe other than BHs. A compact review on the efforts to search for astrophysical wormholes (with an exhaustive list of references) can be found in \cite{Bambi:2021qfo}. Theoretically, the WH can act as a lensing object similarly to the BH and the derivation of the lensing equations would lead to a similar procedure as in the case of the BH \cite{Perlick:2003vg}. Soon after, a plethora of works have been done to distinguish the wormhole from other compact objects with the help of gravitational lensing \cite{chetouani1984geometrical, Clement:1982ej, Perlick:2003vg, Nandi:2006ds, Dey:2008kn, Tsukamoto:2012zz, Kardashev:2006nj, Cramer:1994qj, Harko:2009xf, Tsukamoto:2012xs, Bambi:2013nla}. However, a shred of concrete evidence is still lacking about the existence of wormholes in the universe.
It is commonly believed that the supermassive objects at the galactic centers are black holes; however, their existence depends on the direct detection of the event horizon. Several tests have been performed to detect the event horizon \cite{Broderick:2009ph, Broderick:2015tda, Narayan:2008bv} but the existence of such event horizons is still questionable \cite{Abramowicz:2002vt}. Apart from this, the existence of the shadow boundary does not necessarily give evidence of the existence of the black hole since other compact objects, such as hard spheres \cite{Broderick:2005xa}, naked singularities \cite{Bambi:2008jg, Ortiz:2015rma, Shaikh:2018lcc}, non-rotating wormholes \cite{Bambi:2013nla, Ohgami:2015nra, Ohgami:2016iqm, Mustafa_2021}, and rotating wormholes \cite{Shaikh2018}, can also produce similar shadow boundaries. Such type of work including the possibility of the M87* being a superspinar within the Kerr geometry has been ruled out using the circularity of the image \cite{Bambi:2019tjh} exists in the literature.
Furthermore, after the publication of results on M87$^*$ and Sagittarius A$^*$ (Sgr A$^*$) black hole shadows by Event Horizon Telescope (EHT), interest started to grow for looking into the possibility of wormhole shadows as well \cite{Ohgami:2015nra, Gyulchev:2018fmd, Rahaman:2021web, Ohgami:2016iqm}. In this context, it is worth mentioning that black hole shadows in different modified gravity space-times as well as for compact objects other than black holes, such as naked singularities, hard surfaces, etc., have been studied in great depth in the past (see \cite{Perlick:2021aok, Zakharov:2014lqa, Takahashi:2005hy, Cunha:2015yba, Li:2013jra, Abdujabbarov:2016hnw, Amir:2016cen, Rosa_2022a, Cunha:2016wzk, Atamurotov:2013dpa, Atamurotov:2013sca, Wang:2017hjl, Papnoi:2014aaa, Uniyal:2022xnq, Abdujabbarov:2015rqa, Okyay:2021nnh, Uniyal:2022vdu, Uniyal:2023inx, Rosa_2024, Devi:2021ctm, Roy:2020dyy, Wang:2023nwd, Uniyal:2023ahv, Vagnozzi:2020quf, Roy:2021uye, Ditta:2021uoe, Javed:2022ikp, Mustafa:2022xod, Jusufi:2019knb, Rosa_2022} for a few references). The work has also been done by including plasma space-time around the black hole \cite{Atamurotov:2015nra, Abdujabbarov:2015pqp} as well as for the Kerr de-Sitter background \cite{Chowdhuri:2020ipb}. A lot of work is also available in the literature to understand the bending of light around the compact object (see \cite{Cramer:1994qj, Nandi:2006ds, Nakajima:2012pu, Tsukamoto:2012xs, Tsukamoto:2016zdu, Jusufi:2017mav, Ghosh:2022mka, AbhishekChowdhuri:2023ekr} for a few references) and in the presence of the plasma medium \cite{Adam2015, Bisnovatyi-Kogan:2010flt}.

A plasma medium around the black hole works as a dispersive medium such that light-like geodesics deviate from its path depending on the frequency of the medium, and the geodesic equations can be calculated from the governing Hamiltonian, which can be derived with the help of Maxwell's equations, where the electromagnetic field consists of two charged fluids, with one describing the ions and the other electrons. The transition to ray optics from the known Maxwell's equations in the presence of the plasma within the curved space-time has been given by Breuer and Ehlers \cite{14541db0-3045-3f28-9d31-dc67c5a5c1b8, c5e08cc5-757d-3b1d-a223-fdc681792d68} for the magnetized pressureless plasma. Later on, a similar derivation to calculate the Hamiltonian for the light ray was done by Perlick \cite{perlick2000ray} for the non-magnetized pressureless plasma. Perlick's result for the light-like geodesics can be fully characterized by a direction-independent scalar and refraction, with the latter being a function of the frequency and space-time coordinates. The deflection angle of light has also been discussed for the plasma space-time, where the plasma density is only a function of the radial coordinates in the equatorial plane of Schwarzschild and Kerr BH space-time \cite{perlick2000ray}. A similar result has also been discussed and further analyzed by Bisnovatyi-Kogan and Tsupko \cite{Bisnovatyi-Kogan:2008qbk, Bisnovatyi-Kogan:2010flt, Tsupko:2013cqa}. After that, Morozova et al. \cite{Morozova2013GravitationalLB} generalized the calculation for the off-equatorial plane for the slowly rotating black hole. For more details on how the plasma affects the light ray trajectories, the readers may refer to Er and Mao \cite{Er:2013efa} and Rogers \cite{Rogers:2015dla}.

It is well known that all astrophysical phenomena occur in a plasma-filled spacetime which eventually affects the light trajectories due to it being a dispersive medium \cite{perlick2000ray} in addition to the gravitational influence of compact objects. These plasma distributions do not affect the observation in all frequency ranges, but they have a significant impact on low-frequency, i.e., radio frequency ranges \cite{Muhleman:1970zz, Perlick:2023znh} in which EHT works \cite{EHT2019_1}.
As we know, the plasma medium has an impact on pulsar signal travel time \cite{2004hpa..book.....L}. To explain such behavior, one can assume that the medium is a non-magnetized pressureless plasma and that a linearized theory is sufficient to describe the gravitational field \cite{Perlick_2015}. It is, therefore, interesting to consider the plasma medium and look at the effect of the plasma on such radio frequency ranges. It is safe to assume that black holes as well as exotic compact objects are surrounded by a plasma in a realistic astrophysical setting, and it will therefore be an interesting approach to investigate the observable effects of the plasma, if any, on radio signals that come close to such exotic objects. In case of black holes and compact objects, studies on understanding the imaging and effect of the plasma on deflection of electromagnetic waves (with different space-time metrics) have been performed in detail by Perlick and others; see \cite{Perlick:2017fio, Perlick:2023znh, Tsupko:2013cqa, Liu:2016eju, Rogers:2016xcc, Rogers:2017ofq, Er:2013efa, Bisnovatyi-Kogan:2015dxa, Zhang_2023} and references therein.

In this manuscript, our primary objective is to obtain analytical expressions for the shadow boundary of the Kerr black hole within plasma space-time, particularly for an observer positioned at infinity. We stress that although such a result may seem very trivial and possibly present in the literature, to our surprise, we found that it is not the case. A similar analytical study for high-spin Kerr has been conducted by Yan et al. \cite{Yan_2019}. However, he has assumed incorrect plasma profiles that can't satisfy the separability condition mentioned by Perlick et al. \cite{Perlick:2017fio}.
One of the new results discussed in this paper is the expression for the shadow boundary of the Kerr black hole surrounded by plasma with an observer situated at spatial infinity. 
Because most astrophysical objects are rotating, we compare the rotating BH geometry with a rotating WH geometry. The Kerr geometry is widely accepted with its uniqueness as a rotating axisymmetric solution of Einstein's Field Equations; therefore, we compare the most simple and widely accepted solution for rotating BH, i.e. the Kerr BH with the rotating WH \cite{Teo:1998dp}. Some words are now in order about the rotating WH solutions proposed by \cite{Teo:1998dp}. The rotating Teo WH solution is based on the traversable Morris-Thorne (MT) WH \cite{Morris:1988cz} solution. One of the key features in the MT WH solution is that the conditions that a traversable WH must satisfy is first listed out, and then Einstein equations are used to deduce the form of the matter required to maintain the WH, which is just the opposite of the standard procedure of using the matter content to solve the Einstein's equations. 
It was also shown that the MT WH metric, together with the
wormhole-shaping and traversability conditions, imply that the corresponding stress-energy tensor violates the null, and therefore also the weak, energy condition \cite{Visser1996}. Such a matter content was termed as `exotic' which in turn showed that it is almost impossible to do interstellar travels using WHs. However, not everything was lost, and it was shown \cite{PhysRevD.39.3182} that by giving up on spherical symmetry, it is possible for an observer falling through the wormhole to avoid encountering the exotic matter. Motivated by this, Teo \cite{Teo:1998dp} has proposed the stationary, axisymmetric generalization of the MT WH. Although the null energy condition is generically violated at the throat in such a WH, but it is possible for geodesics falling through such WHs to avoid the energy-condition violating matter and thereby avoiding any problems related to that.
In this work, therefore, we seek to conduct a comparative investigation between the shadows of the Kerr black hole and that of the wormhole in a homogeneous plasma space-time, provided the plasma distribution satisfies the Hamiltonian separability criterion \cite{Perlick:2017fio, Bezdekova:2022gib, kumar2023shadow}. 

Another essential aspect of our study is the derivation of the deflection angle for light in the vicinity of the Kerr black hole, considering both homogeneous and non-homogeneous plasma space-time. By analyzing their impacts on the deflection angle and comparing them with the deflection angle of the rotating wormhole, we aim to differentiate the distinct characteristics of these two astrophysical objects. A thorough analysis of shadow and weak lensing by a rotating wormhole in plasma spacetime has been discussed by Kumar et al. \cite{kumar2023shadow} and our comparison with the wormhole is based on the calculations presented in the said work. Furthermore, our research endeavors to constrain the plasma parameters, utilizing observational data from the Event Horizon Telescope (EHT) of black hole shadows at the core of M87* \cite{EventHorizonTelescope:2019dse,PhysRevD.100.044057}. This part of our research builds upon similar approaches and calculations as previously discussed in \cite{Rahaman_2021}.

The paper is organized as follows: Section II presents a brief overview of the Hamiltonian formalism for null geodesics in plasma space-time. Moving on to Section III, we derive the precise expressions for the null geodesic equations, focusing on specific plasma profiles that satisfy the crucial condition for the existence of Carter's constant. In Section IV, our focus shifts to the celestial coordinates of the shadow boundary, where we conduct a thorough comparison of shadows for varying plasma densities of multiple plasma distributions. Additionally, this section explores the intriguing contrast between the shadows of the Kerr black hole and the Teo wormhole in a homogeneous plasma space-time. Section V delves into the calculations of the deflection angle by a slow-rotating black hole, employing the weak field approximation for both uniform and non-uniform plasma space-times. The results are compared with those obtained for the rotating wormhole \cite{kumar2023shadow}. Lastly, Section VI is dedicated to the effort of constraining the plasma parameters, taking into account the observational data gathered from M87$^*$. Throughout the paper, we have adopted units, setting $\hbar=G=c=m=1$, and employed a signature of (-,+,+,+).

\section{HAMILTON FORMALISM FOR LIGHT RAYS IN KERR BLACK HOLE PLASMA space-time and Teo wormhole metric}
It is well known that a rotating, stationary, axisymmetric, and asymptotically flat black hole solution is described by the Kerr metric in the Boyer-Lindquist coordinates as
\begin{align}
\begin{split}
    d s^2=-\left(1-\frac{2 m r}{\rho^2}\right)d t^2-\left(\frac{4 m r a\sin^2{\theta}}{\rho^2}\right)d t d\phi \\+\left(\frac{\rho^2}{\Delta}\right)d r^2   +\rho^2 d\theta^2\\+  \left(r^2+a^2+\frac{2m r a^2\sin^2{\theta}} {\rho^2}\right) \sin^2{\theta} d\phi^2,
\end{split}
\label{eq2.1.1}
\end{align} 
where
\begin{align}
\label{eq2.1.2}
\begin{split}
    \rho^2 &=r^2+a^2\cos^2{\theta},\\
    \Delta&=r^2-2mr+a^2,
\end{split}
\end{align}
and $a=J/m$ with $a^2\leq m^2$ \cite{Frolov1998}, $m$ and $J$ are known as the mass and angular momentum of the Kerr black hole respectively. Now, considering the Hamiltonian in the presence of the plasma density for the light ray traveling in non-magnetized pressure-less plasma as
\begin{equation}
    H(x^\alpha, p_\mu)=\frac{1}{2}\left(g^{\mu \nu}(x^\alpha) p_\mu p_\nu+\omega_e(x^\alpha)^2\right),
\label{2.3}
\end{equation}
where $g^{\mu \nu}$ are the contravariant components of the metric function and $\omega_e$ represents the plasma electron frequency given by,
\begin{equation}
    \omega_e(x^\alpha)^2=\frac{4 \pi e^2}{m_e} N(x^\alpha), 
\end{equation}
where $m_e$ and $e$ are the mass and charge of the electron, respectively while $N(x^\alpha)$ defines the electron density. The plasma frequency ($\omega_e$) and the photon frequency ($\omega$) are related by a general form,
\begin{equation}
    n(x^\alpha, \omega(x^\alpha))^2=1-\frac{\omega_e(x^\alpha)^2}{\omega(x^\alpha)^2},
\end{equation}
where $n$ is known as the refractive index.  Considering the gravitational redshift for the constant of motion $p_t$, the observed redshift frequency is expressed as,
\begin{equation}
    \omega(x^\alpha)=\frac{p_t}{\sqrt{-g_{tt}(x^\alpha)}}.
\label{6}
\end{equation}
Considering the Hamiltonian expression (Eq. \ref{2.3}), the geodesic equations of light rays can be derived using Hamilton's equations
\begin{equation}
    \dot{p}_\mu=-\frac{\partial H}{\partial x^\mu}, ~ \dot{x}^\mu=\frac{\partial H}{\partial p_\mu}.
\label{7}
\end{equation}
Since we are doing our analysis in the context of plasma space-time and light rays interacting with the plasma medium (we have considered here only the dispersive nature of plasma and neglected its gravitational effect), it becomes necessary to find out if light reaches the observer or not. The necessary and sufficient condition for the existence of a light ray with the constant of motion $p_t$ has been derived for the Kerr black hole in  \cite{Perlick:2017fio} and is given by 
\begin{equation}
    p_t^2 > \left(1-\frac{2mr}{\rho^2}\right) \omega_e(x^\alpha)^2.
\label{8}
\end{equation}
This criterion should satisfy the condition of outer communication, i.e., it will pave the path for the light rays to reach the observer.

In Secs. IV and V, we have made some analytical comparison of the Kerr black hole with the wormhole, and for our analysis, we have considered a stationary, axisymmetric rotating metric for  Teo class traversable wormhole in the Boyer-Lindquist coordinates as \cite{Teo1998},
\begin{align}
\begin{split}
ds^2 = & -N(r)^2 dt^2 + \left(1 - \frac{b_0(r)}{r}\right)^{-1} dr^2 \\
& + r^2 K(r)^2 \left(d\theta^2 + \sin^2\theta \left(d\phi - \omega_T(r) dt\right)^2\right),
\end{split}
\end{align}
where $N$, $b_0$, $K$, and $\omega_T$ are given as
\begin{equation}
\begin{aligned}
& N=\exp \left[-\frac{r_0}{r}\right], \quad b_0(r)=r_0=2M, \\
& K=1, \quad \omega_T=\frac{2 J}{r^3},
\end{aligned}
\end{equation}
where $r \geq r_0 $, $r_0$ is the wormhole throat radius, $J$ is the angular momentum of the wormhole and $M$ is the mass of the wormhole \cite{Shaikh2018, Visser1996}. See \cite{Teo1998, kumar2023shadow} for more details on this wormhole geometry.

\section{NULL GEODESIC EQUATIONS IN PLASMA ON KERR BLACK HOLE space-time}
In this section, we understand the intricacies of the Hamilton-separable condition and its implications for geodesic equations in plasma space-time. When studying the motion of particles in this context, we encounter two well-known constants of motion: the angular momentum of the particle around the axis of symmetry ($p_{\phi}$), and its energy ($p_t$). However, an additional hidden constant exists, known as Carter's constant \cite{carter1968global}.

Carter's constant emerges as a consequence of the Hamilton-separable condition, which governs the motion of light rays in the latitudinal direction. By satisfying this condition, we can achieve a separable Hamiltonian, an essential requirement for generalized plasma contributions. The Hamiltonian for null geodesics, given by $H\left(x^\alpha,p_\mu \right)=0$, holds a crucial role in capturing the dynamics of these particles, and we aim to derive it in the context of plasma space-time. Therefore, since the plasma frequency depends on both the radial coordinate ($r$) and the polar angle ($\theta$), the Hamiltonian equation presented can only be separable if we consider the general form of the plasma frequency as \cite{Perlick:2017fio},
\begin{equation}
    \omega _e (r,\theta)^2 =\frac{\Omega _r(r)+\Omega _{\theta}(\theta)}{r^2+a^2 \cos^2{\theta}}.
\label{3.9}
\end{equation}
By taking this generalized form of the plasma frequency, we ensure that the equation exhibits the desired separability properties. This allows us to analyze the impact of plasma on the dynamics of light rays, considering both its radial and latitudinal variations. On simplifying Eq. (\ref{2.3}) for null geodesic orbits using Eq. (\ref{3.9}), we get
\begin{equation}
\begin{split}
    \underbrace{\frac {-\left(p_t(r^2+a^2)+ a  p_{\phi}  \right)^2 +  \Delta^2 p_{r}^{2}}{\Delta} +\Omega_r(r)}_{f_r(r)}\\ =
     \underbrace{-\Bigg[a p_t \sin{\theta} - \frac{p_{\phi}}{\sin{\theta}}\Bigg]^2 - p_{\theta}^{2}-\Omega_{\theta}(\theta)}_{f_{\theta}(\theta)}.
\label{3.10}
\end{split}
\end{equation}
Since expressions in the LHS and RHS of Eq. (\ref{3.10}) are dependent on $r$ and $\theta$ coordinates, respectively, the equality tells us that both the expressions $f_r(r)$ and $f_{\theta}(\theta)$ are constant. Therefore, we can define them as
\begin{equation} 
f_r(r)=f_{\theta}(\theta) = - \kappa.
\end{equation}

This newly introduced constant, denoted as $\kappa$, serves as a generalization of Carter's constant \cite{carter1968global}. It plays a crucial role in characterizing the motion of massless particles in the $\theta$ direction, which is necessary for determining the properties of spherical photon orbits, photon regions, and ultimately, the shadow of a Kerr black hole.

Previously, we had already discussed two conserved quantities along the paths of photons: $p_t$ and $p_{\phi}$. With the addition of Carter's constant ($\kappa$), we now possess three constants that significantly impact the equations governing the geodesic motion of photons. However, these three constants can be effectively reduced to two impact parameters.
\begin{equation}
    \eta=\frac{L}{\omega_o}, ~ \xi=\frac{\kappa}{\omega_o^2},
\label{3.12}
\end{equation}
where we have assumed $p_t=-\omega_o$ and $p_\phi=L$.
On solving Hamilton's equations (\ref{7}) for $x^\mu=t, \phi$, we get
\begin{equation}
    \dot{t}= \frac{1}{\Delta}\Bigg(\left( r^2+a^2+\frac{2m r a^2 \sin^2{\theta}}{\rho^2}
    \right)  -\frac{2mra}{\rho^2}\eta \Bigg),
\end{equation}
\begin{equation}
    \dot{\phi}= \frac{1}{\Delta \rho^2}\Bigg(\frac{\rho^2-2mr}{\sin^2{\theta}}\eta +2mra\Bigg).
\end{equation}

While the expressions for $p_r$ and $p_\theta$ are derived by using impact parameters and Eq. \ref{3.10} as,
\begin{equation}
   p_r= \pm \frac{1}{\Delta}\sqrt{((r^2+a^2-a\eta)^2-\Delta(\xi+\frac{\Omega_r}{\omega_o^2})},
\end{equation}
\begin{equation}
     p_{\theta} = \pm\sqrt{\xi - \frac{\Omega_{\theta}}{\omega_o^2} - \Bigg( a \sin{\theta}-\frac{\eta}{\sin{\theta}} \Bigg)^2},
\end{equation}
and using the expressions for $p_r$ and $p_\theta$ we solve Hamilton's equations (\ref{7}) for $x^\mu=r, \theta$ and given as,
\begin{equation}
    \dot{r} = \pm \frac{1}{\rho^2} \sqrt{R(r)},
\label{3.17}
\end{equation}
\begin{equation}
    \dot{\theta} = \pm \frac{1}{\rho^2} \sqrt{ \Theta(\theta)},
\end{equation}
where the functions $R(r)$ and $\Theta(\theta)$ are expressed by,
\begin{equation}
    R(r) = (r^2+a^2-a  \eta)^2 - \Delta\left(\xi+\frac{\Omega_r}{\omega_o^2}\right),
\end{equation}
\begin{equation}
    \Theta(\theta)= \xi - \frac{\Omega_{\theta}}{\omega_o^2} - \Bigg( a  \sin{\theta}-\frac{\eta}{\sin{\theta}} \Bigg)^2,
\end{equation}
and to study unstable spherical orbits, one must ensure that the following conditions are satisfied:
\begin{equation}
    \Theta(\theta)\geq 0 ~,~~R''(r) >0.
\label{21}
\end{equation}
Please note that the dot indicates a derivative with respect to an affine parameter, and the prime indicates a derivative with respect to a radial coordinate.
\begin{figure*}
\subfigure(a){\includegraphics[width = 2.9in, height=2.7in]{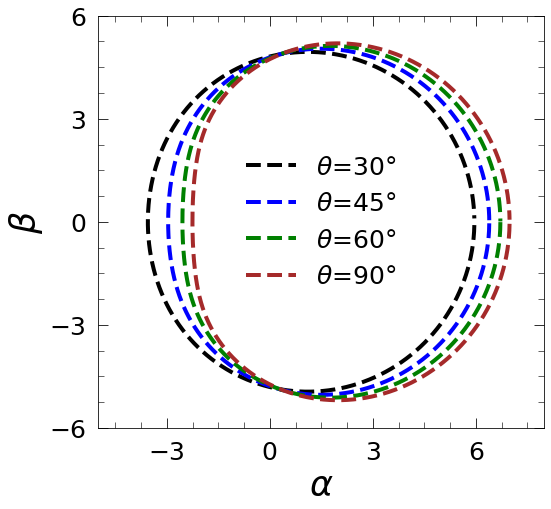}}
\subfigure(b){\includegraphics[width = 2.9in, height=2.7in]{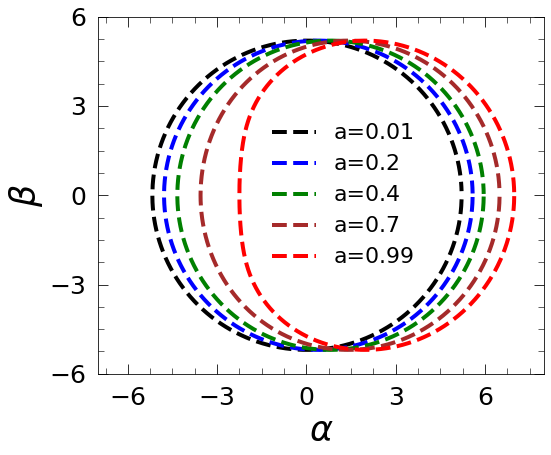}}
\caption{Kerr black hole shadows in vacuum for (a) various inclinations angles with $a=0.99$ and (b) various spin parameters with $\theta=90^\circ$.}
\label{fig:1}
\end{figure*}
\section{SHADOW OF KERR BLACKHOLE IN PLASMA space-time}
In this section, we delve into the study of plasma distribution profiles in the vicinity of black holes. Shapiro et al. \cite{shapiro1974accretion} conducted pioneering research on accretion onto black holes, revealing that the plasma frequency follows a proportionality to $r^{-3/2}$ for pressureless non-magnetized plasma. However, for our analytical model, we are considering generalized plasma distribution models while adhering to the separability condition given by Eq. \ref{3.9}.

To facilitate our analysis, we shall first determine the general form of celestial coordinates for the last photon rings. These photon orbits provide valuable insights into the optical appearance of a Kerr black hole and contribute to our understanding of its characteristics and behavior. Our primary objective is to identify the specific photon orbits that distinguish the boundary between ingoing and outgoing light rays. To achieve this, we utilize the impact parameters $\eta$ and $\xi$, which play a crucial role in characterizing these critical orbits and, consequently, determining the boundary of the shadow region. It is worth noting that these critical orbits correspond to the most unstable circular trajectories, with the maximum value of the effective potential $V_{eff}$. The conditions that determine the existence of these unstable circular photon orbits have been extensively studied in the literature \cite{Teo2003, Frolov1998} and these conditions provide us with a set of standard criteria to identify and analyze these orbits and are given as,
\begin{equation}
    \left.V_{eff}(r)\right|_{r=r_c}=0, \quad \left.{V{'}_{eff}(r)}\right|_{r=r_c}=0,
\label{22}
\end{equation}
where $r_c$ denotes the critical photon orbits. Since the geodesic equation in the radial direction (Eq. \ref{3.17}) can be written as $\dot{r}^2/2+V_{eff}=0$ hence the conditions (Eq. \ref{22}) can be modified as,
\begin{equation}
    \left.R(r)\right|_{r=r_c} =0, ~~ \left.{R{'}(r)}\right|_{r=r_c} =0,
\label{23}
\end{equation}
and solving for these conditions we get the expressions for impact parameters such as,
 \begin{equation}
    \xi = \left.\bigg[\frac{1}{\Delta}(r^2+a^2-a\eta)^2-\frac{\Omega_r}{\omega_o^2}\Bigg]\right|_{r=r_{c}},
\end{equation}
 \begin{equation}
     \eta = \frac{B-\sqrt{B^2-4AC}}{2A},
 \end{equation}
where,
\begin{align}
    A &=  \frac{\Delta'}{\Delta}a^2,  \\
    B &=   \bigg(\frac{\Delta'}{\Delta}2a(r^2+a^2)-4ar\Bigg),  \\
    C &=  -4(r^3+ra^2)+\frac{\Delta'}{\Delta}(r^2+a^2)^2+\frac{\Omega_r'}{\omega_o^2}\Delta.
\end{align}
It is important to understand that $\eta$ and $\xi$ can completely determine the boundary of the shadow but to form the shadow in the observer's sky, we need to find out the celestial coordinates, which are defined as \cite{Vázquez_Esteban_2004},
\begin{align}
    \alpha &=\lim _{r \rightarrow \infty}\left(-r^2 \sin \theta \frac{d \phi}{d r}\right), \label{29}\\
    \beta &=\lim _{r \rightarrow \infty} r^2 \frac{d \theta}{d r} \label{30},
\end{align}
therefore with the help of geodesic equations, we can write the expression for the celestial coordinates as,
\begin{align}
    \alpha &= -\frac{\eta}{ \sin{\theta}},   \label{31}\\
    \beta &= \sqrt{\xi-(\eta \csc \theta-a \sin \theta)^2- \frac{\Omega_{\theta}}{\omega_o^2}} \label{32}.
\end{align}

However, it is essential to note that these expressions do not hold in the case of a homogeneous plasma distribution. Considering the vacuum case, Fig. \ref{fig:1} (a) and Fig. \ref{fig:1} (b) display the shadows of a Kerr black hole for different inclination angles and varying spin parameters respectively. It is evident from the figures that as the inclination angle and spin parameter increase, the shape of the shadow deviates from circularity. However, the spin parameter significantly changes the shape of the circularity of the shadow. This highlights the impact of the black hole's rotation on the observed shadow, leading to distortions distinct from the circular shape. However, this does not take care of the plasma medium; therefore, in the subsequent sections, we will try to understand the effect of the plasma medium on the shadow of the black hole.

\subsection{Homogeneous plasma space-time}
\begin{figure*}
\subfigure{(a)}{\includegraphics[width = 2.16in, height=2in]{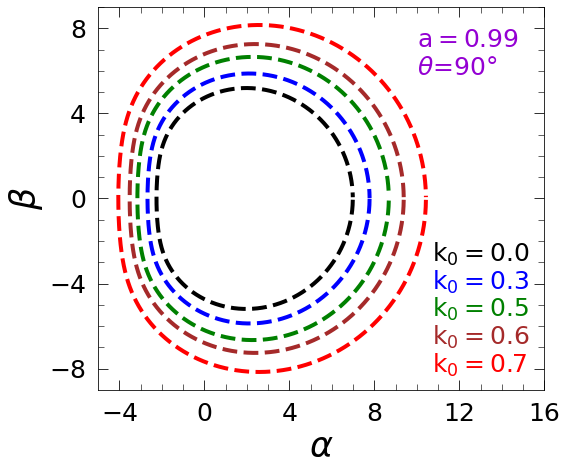}}
\subfigure{(b)}{\includegraphics[width = 2.16in, height=2in]{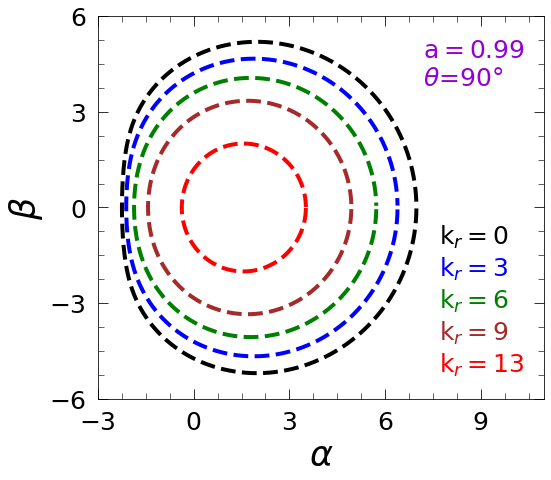}}
\subfigure{(c)}{\includegraphics[width = 2.16in, height=2in]{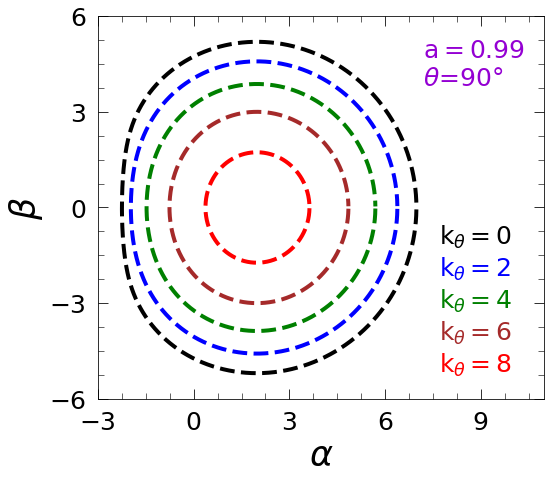}}
\caption{Comparison of Kerr black hole shadows with various plasma parameters ($k_0$, $k_r$, $k_\theta$) for plasma distributions: (a) $\omega_e^2=k_0\omega_o^2$, (b) $\omega_e^2=\frac{k_r\sqrt{r}}{r^2+a^2\cos^2\theta}\omega_o^2$, and (c) $\omega_e^2=k_\theta\frac{1+2\sin^2\theta}{r^2+a^2\cos^2\theta}\omega_o^2$ .}
\label{fig:2}
\end{figure*}

Here, we will assume an ideal situation for plasma distribution, which is homogeneous plasma distribution and is generally given by:
\begin{equation}
    \frac{\omega_e^2}{\omega_o^2}=k_0,
\label{33}
\end{equation}
where $k_0$ denotes the homogeneous plasma parameter and should be in the range $(0,1)$ to satisfy Eq. \ref{8}. Now, using Eqs. \ref{3.9} and \ref{33}, we get the following expressions:
\begin{equation}
\Omega_r=k_0r^2\omega_o^2 ~,~ ~ \Omega_{\theta}=k_0a^2\cos^2\theta \omega_o^2.
\end{equation}
As we have mentioned earlier, the shadow boundaries for homogeneous plasma profile are not given by Eqs. \ref{31} and \ref{32}, so the celestial coordinates in this case are derived by solving Eqs. \ref{29} and \ref{30} along with geodesic equations and is given by,
\begin{align}
\begin{split}
    \alpha &= -\frac{\eta  \csc \theta }{\sqrt{1-k_0}},  \\
    \beta &=\frac{\sqrt{\xi-k_0 a^2 \cos ^2\theta -(\eta  \csc \theta -a \sin \theta )^2 }}{\sqrt{1-k_0}}.
\end{split}
\end{align}
In contrast to the plasma distribution as $r^{-3/2}$ \cite{shapiro1974accretion} (one of the models for non-uniform plasma distributions), which approaches zero density as the radial distance tends to infinity ($r \rightarrow \infty$), the homogeneous plasma introduces a non-zero density throughout the space-time. Particularly, in the case of a uniform distribution, the plasma density plays a crucial role in shaping the shadow of a black hole. Fig. \ref{fig:2} (a) demonstrates the increasing impact of this density on the black hole shadow, emphasizing the non-uniform distribution of plasma in the vicinity of the black hole since the shadow size in a homogeneous plasma space-time is larger and would have detected with low-resolution radio telescopes. 

\subsection{Plasma distribution with \texorpdfstring{$\Omega_{\theta}=0$}{}}
In this particular case, we assume such plasma profile, where  $\Omega_{\theta}=0$. This condition, specified by the plasma profile condition (Eq. \ref{3.9}), restricts us from selecting a radial plasma profile independent of $\theta$, as mentioned previously. To emulate the behavior of $r^{-1.5}$, a corresponding profile, as proposed by Perlick et al. \cite{Perlick:2017fio}, is adopted and expressed as follows:
\begin{equation}
    \frac{\omega_e^2}{\omega_o^2}=\frac{k_r\sqrt{r}}{r^2+a^2\cos^2\theta},
\label{36}
\end{equation}
and consequently we can get the expression for $\Omega_r$ and $\Omega_\theta$ with the help of Eqs. \ref{3.9} and \ref{36} as,
\begin{equation}
    \Omega_r=k_r\sqrt{r}\omega_o^2 ~,~ ~ \Omega_{\theta}=0,
\end{equation}
and the celestial coordinates for this plasma profile using Eqs. \ref{31} and \ref{32} are given by,
\begin{equation}
    \alpha = -\eta  \csc \theta ~,~~ \beta=\sqrt{\xi -(\eta  \csc \theta -a \sin \theta )^2}.
\end{equation}
In contrast to the uniform distribution, the plasma profile under consideration exhibits a negative impact on the black hole shadow, as illustrated in Fig. \ref{fig:2} (b) for spin parameter $a=0.99$ and inclination angle $\theta=90^\circ$. In this case, the radial plasma parameter ($k_r$) is determined in such a manner that it satisfies the conditions for unstable circular orbits given by Eqs. \ref{21} and \ref{23} and also ensures the presence of light in the outer communication region of the Kerr black hole given by Eq. \ref{8}. It is worth noticing that as we increase the radial plasma parameter, the shadow size decreases, and hence there exists an upper limit for the plasma density for which the shadow vanishes completely, which can be easily inferred from Fig. \ref{fig:2} (b).

\begin{figure*}
\subfigure{\includegraphics[width = 2.25in, height=2.4in]{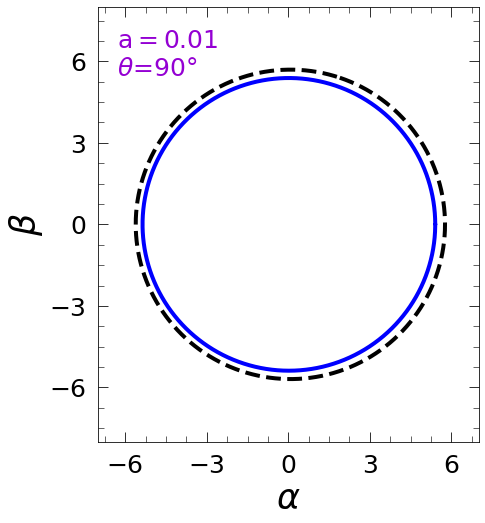}}
0\subfigure{\includegraphics[width =  2.25in, height=2.4in]{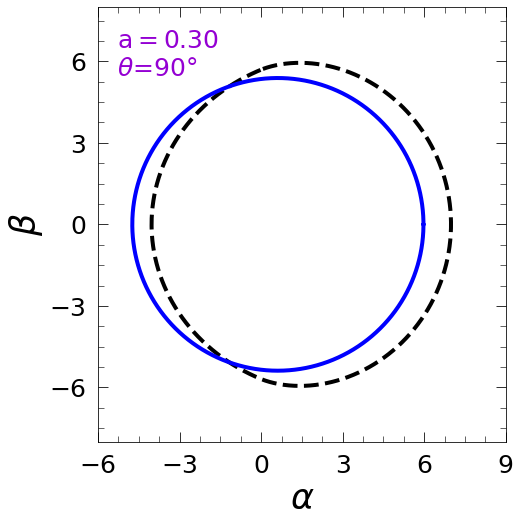}}
\subfigure{\includegraphics[width =  2.25in, height=2.4in]{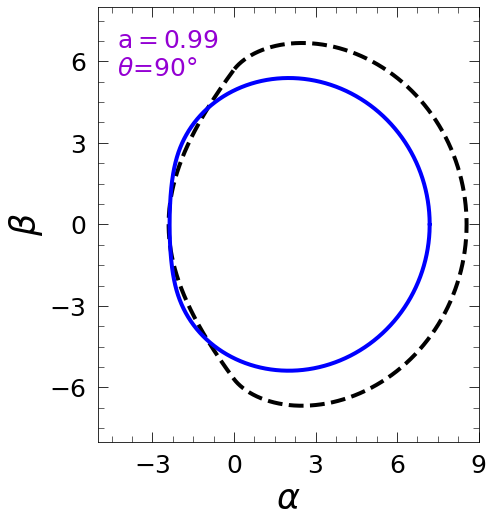}}
\subfigure{\includegraphics[width = 2.25in, height=2.4in]{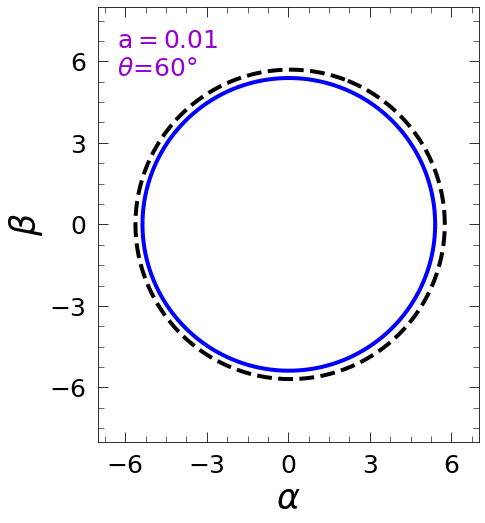}}
\subfigure{\includegraphics[width = 2.25in, height=2.4in]{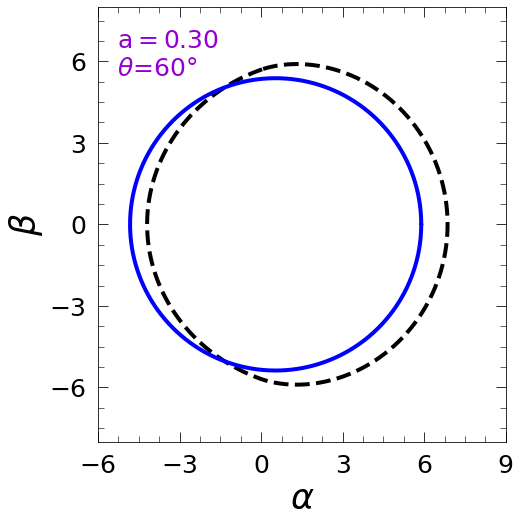}}
\subfigure{\includegraphics[width = 2.25in, height=2.4in]{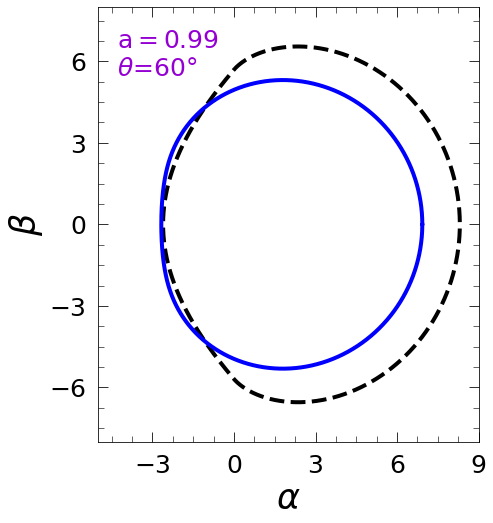}}
\subfigure{\includegraphics[width = 2.25in, height=2.4in]{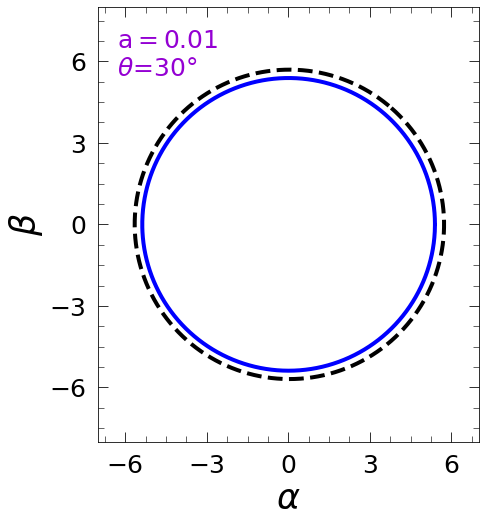}}
\subfigure{\includegraphics[width = 2.25in, height=2.4in]{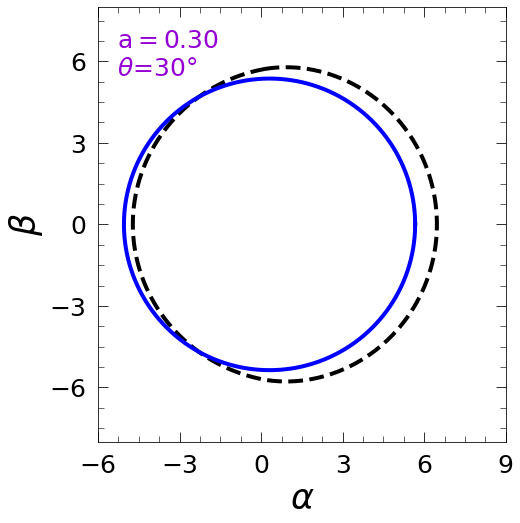}}
\subfigure{\includegraphics[width = 2.25in, height=2.4in]{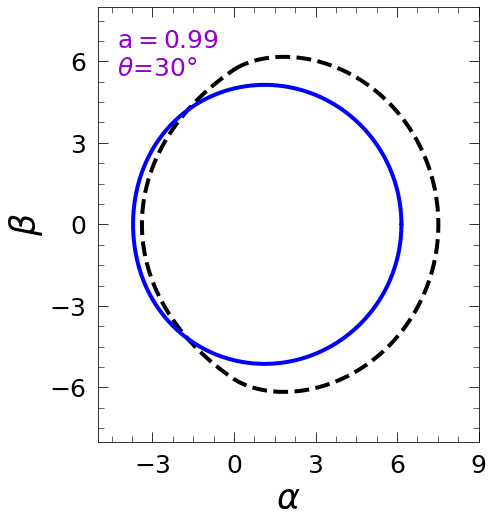}}

\caption{The shadows of Kerr black hole (solid blue curve) and the shadows of Wormhole (dashed black curve) with homogeneous plasma parameter, $k_0= 0.1$. Comparison has been made by considering homogeneous plasma profile as $\omega_e^2=k_0\omega_o^2$.}
\label{fig:3}
\end{figure*}

\subsection{Plasma distribution with \texorpdfstring{$\Omega_r=0$}{}}
In this part, we consider an additional scenario where the plasma distribution exhibits dependence on the $\theta$ coordinate, thus $\Omega_r=0$. The Hamilton separable condition (Eq. \ref{3.9}) restricts us from selecting a plasma profile entirely dependent on $\theta$. A specific plasma profile, as proposed by Perlick et al. \cite{Perlick:2017fio}, is adopted and given as follows:
\begin{equation}
    \frac{\omega_e^2}{\omega_o^2}=k_\theta\frac{1+2\sin^2\theta}{r^2+a^2\cos^2\theta},
\label{39}
\end{equation}
and using Eqs. \ref{3.9} and \ref{39}, we can write the following expressions:
\begin{equation}
    \Omega_r=0 ~,~~ \Omega_{\theta}=k_\theta(1+2\sin^2\theta) \omega_o^2,
\end{equation}
Here, $k_\theta$ is the latitudinal plasma parameter, and its values are determined using Eqs. \ref{8} and \ref{21}. The celestial coordinates for this case can be expressed using Eqs. \ref{31} and \ref{32} as,
\begin{align}
\begin{split}
    \alpha &= -\eta  \csc \theta,  \\
\beta &= \sqrt{\xi -(\eta  \csc \theta -a \sin \theta)^2-k_\theta(1+2 \sin ^2\theta)}.
\end{split}
\end{align}
Fig. \ref{fig:2} (c) shows the comparative shadow of the black hole for this particular case of plasma profile. It can be easily inferred that the only dependency on $\theta$ for plasma density has a similar effect on the shadow as compared to the previously mentioned profile given by Eq. \ref{36}.
As one can notice, the numerator in the previous plasma profile includes $\Omega(r)$ while this profile includes $\Omega(\theta)$, and these differences show a significant impact on the BH shadow as shown in Fig. \ref{fig:2}. As we can conclude from Figs. \ref{fig:2} (b) and \ref{fig:2} (c), the shadow radius shrinks more in the case of this plasma profile compared to the previous profile as plasma parameters increase, i.e., the shadow radius is similar for both profiles at $k_\theta$ less than $k_r$. This finding tells us that we should include the latitudinal dependency of plasma distribution in the low-frequency observational study of compact objects.

Given the distinctive plasma profile requirements for the Teo wormhole \cite{kumar2023shadow}, contrasting those of the Kerr black hole given by Eq. \ref{3.9}, we have exclusively examined the comparison between the rotating wormhole and the rotating black hole for a homogeneous plasma profile; otherwise, if we had chosen a different plasma profile, it would not be possible to find Carter's constant, i.e., Hamilton-Jacobi variable separation wouldn't be performed since our aim here is to compare the shadow boundaries of both compact objects analytically. Fig. \ref{fig:3} depicts the comparison of the shadow profiles for the wormhole and black hole, considering a plasma parameter of $k_0=0.1$ along with varying spin parameters and inclination angles. Notably, the comparison reveals remarkable shadow contour size similarity for lower spin parameters. However, at higher spin parameters, the shadow of the Teo wormhole exhibits a greater deviation from circularity compared to that of the Kerr black hole.

\section{Weak gravitational lensing}

In this section, we investigate the influence of plasma distributions on the deflection angle within the framework of weak field approximation. As we know, when light rays traverse in the vicinity of massive objects, they undergo deviations from their original trajectories. Furthermore, the presence of plasma introduces additional complexity in such a way that it can either positively or negatively affect the deflection angle, depending on the particular plasma distribution. Notably, all calculations presented herein are performed for a specific case where the observer is situated at infinity and in the equatorial plane of the source, which corresponds to $\theta=90^\circ$.

The calculation of the deflection angle involves analyzing the deviation of light from its initial trajectory. Thus, at the distance of the closest approach (R), the condition to calculate the deflection angle can be expressed as follows,
\begin{equation}
    \left.\left(\frac{\dot{r}}{\dot{\phi}}\right)\right|_{r=R}=0.
\end{equation}
Now, for calculating the deflection angle by Kerr black hole in the plasma space-time, we have used the generalized analytic expression for the deflection angle by rotating space-time as derived by Kumar et. al \cite{kumar2023shadow} and with the help of expression given by Eq. 65 from \cite{kumar2023shadow}, the impact parameter ($\lambda = \omega_o/L$) can be written as,
\begin{equation}
\lambda=\left.\frac{2 g ^{\phi t} \pm \sqrt{(2 g ^{\phi t})^2-4 g ^{\phi \phi}\left(g^{t t}+\omega_e^2/\omega_o^2\right)}}{2\left(g^{t t}+\omega_e^2/\omega_o^2\right)}\right|_{r=R},
\end{equation}
while the integral form of the deflection angle of light from its original trajectory is given by Eq.66 from \cite{kumar2023shadow} and is expressed as,
\begin{widetext}
\begin{equation}
 \int_0^{\bar{\alpha}} d \phi=\pm 2 \int_R^{\infty}\left[\frac{-g^{r r}}{\left(g^{\phi \phi}-g^{\phi t} \lambda\right)^2}\left((g^{t t}+\omega_e^2/\omega_o^2) \lambda^2-2 g^{\phi t} \lambda+g^{\phi \phi}\right)\right]^{-1 / 2} d r.
 \label{44}
\end{equation}
\end{widetext}
Here, we have considered the integral from R to $\infty$ instead of $-\infty$ to $\infty$ since the deflection is symmetric about the closest distance of approach (R). The final deflection angle from the original trajectory is given by $\alpha=\bar{\alpha}-\pi$ since the choice of coordinate is such that the massive object is at the origin and if there were no massive objects then the deviation of light would be $\pi$. To account for the symmetry of deflection around the closest approach point, the integral expression given by Eq. 66 from \cite{kumar2023shadow} is adjusted, with integration now performed from R to $\infty$.

In our analysis, we calculate deflection angles for two specific cases one where plasma is considered homogeneous and another when plasma is radially distributed in space-time. Homogeneous plasma features uniform density while radial plasma exhibits density variation along the radial direction. Examining these cases enhances our understanding of the influence of plasma on light deflection and gravitational lensing phenomena. We consider the plasma distributions described in \cite{Bisnovatyi-Kogan_2010, farruh2021} in the following subsections.
\begin{figure*}
\subfigure(a){\includegraphics[width = 3in, height=2.7in]{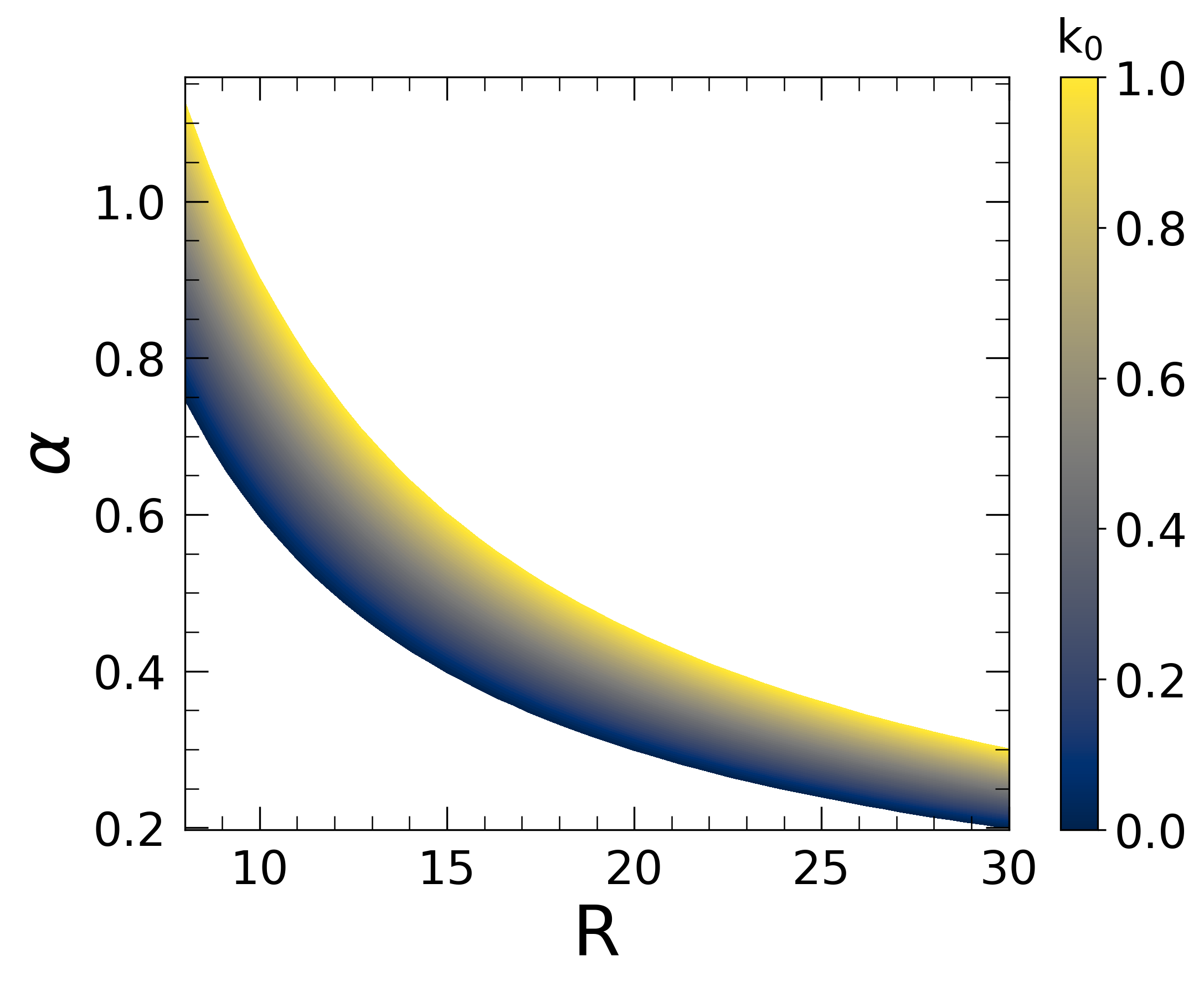}}
\subfigure(b){\includegraphics[width = 3in, height=2.7in]{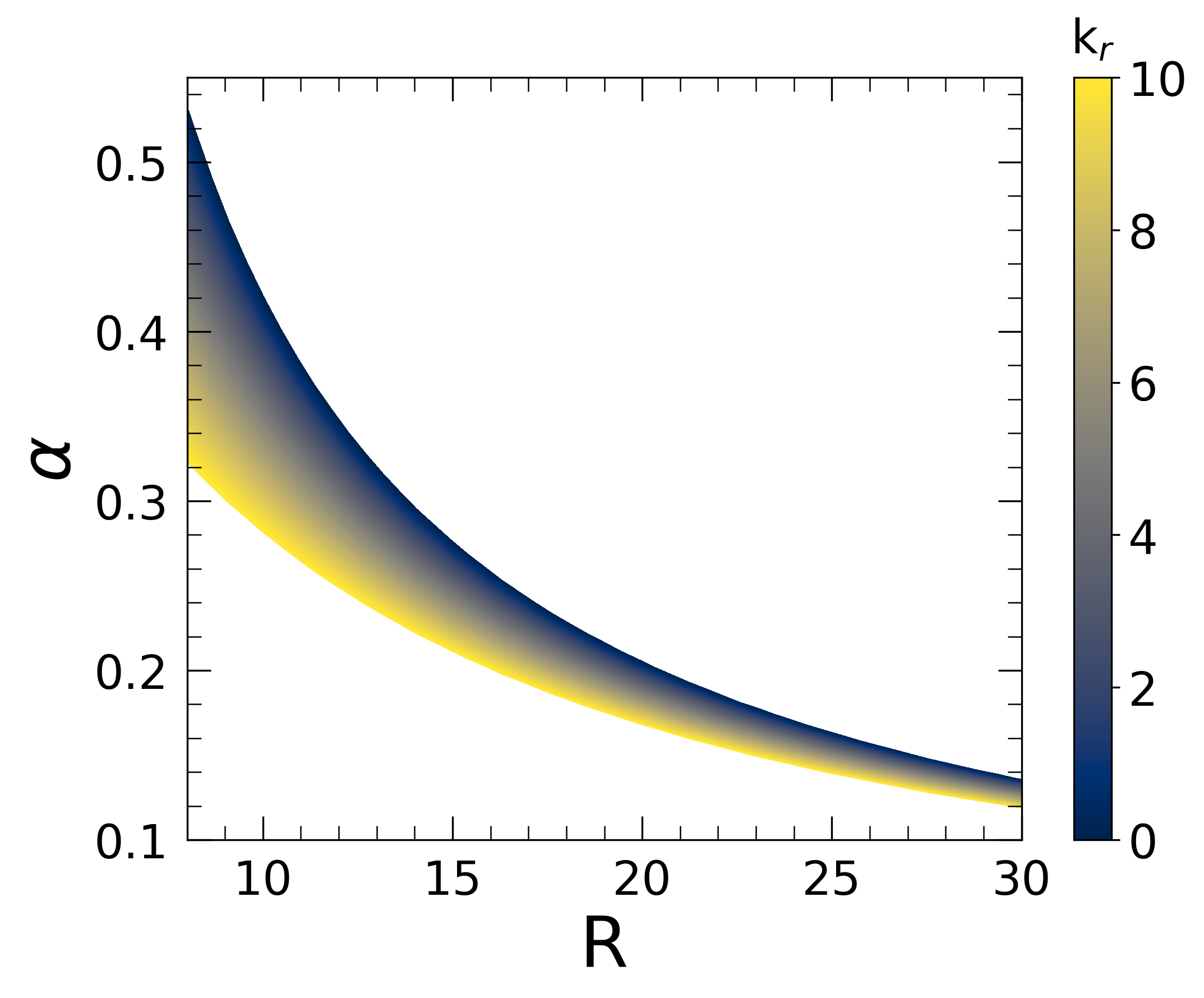}}
\caption{Weak deflection angle by Kerr black hole as a function of closest distance approach (R) in plasma space-time for various plasma parameters ($k_0$, $k_r$) of plasma profiles (a) $\omega_e^2=k_0\omega_o^2$, Here the plasma parameter is $k_0$, and (b) $\omega_e^2=\frac{k_r}{r^2}\omega_o^2$, here the plasma parameter is $k_r$ and its value is taken accordingly to satisfy the low-plasma density and Eq. \ref{8}.}
\label{fig:4}
\end{figure*}
\begin{figure*}
\subfigure(a){\includegraphics[width = 3.0in, height=2.7in]{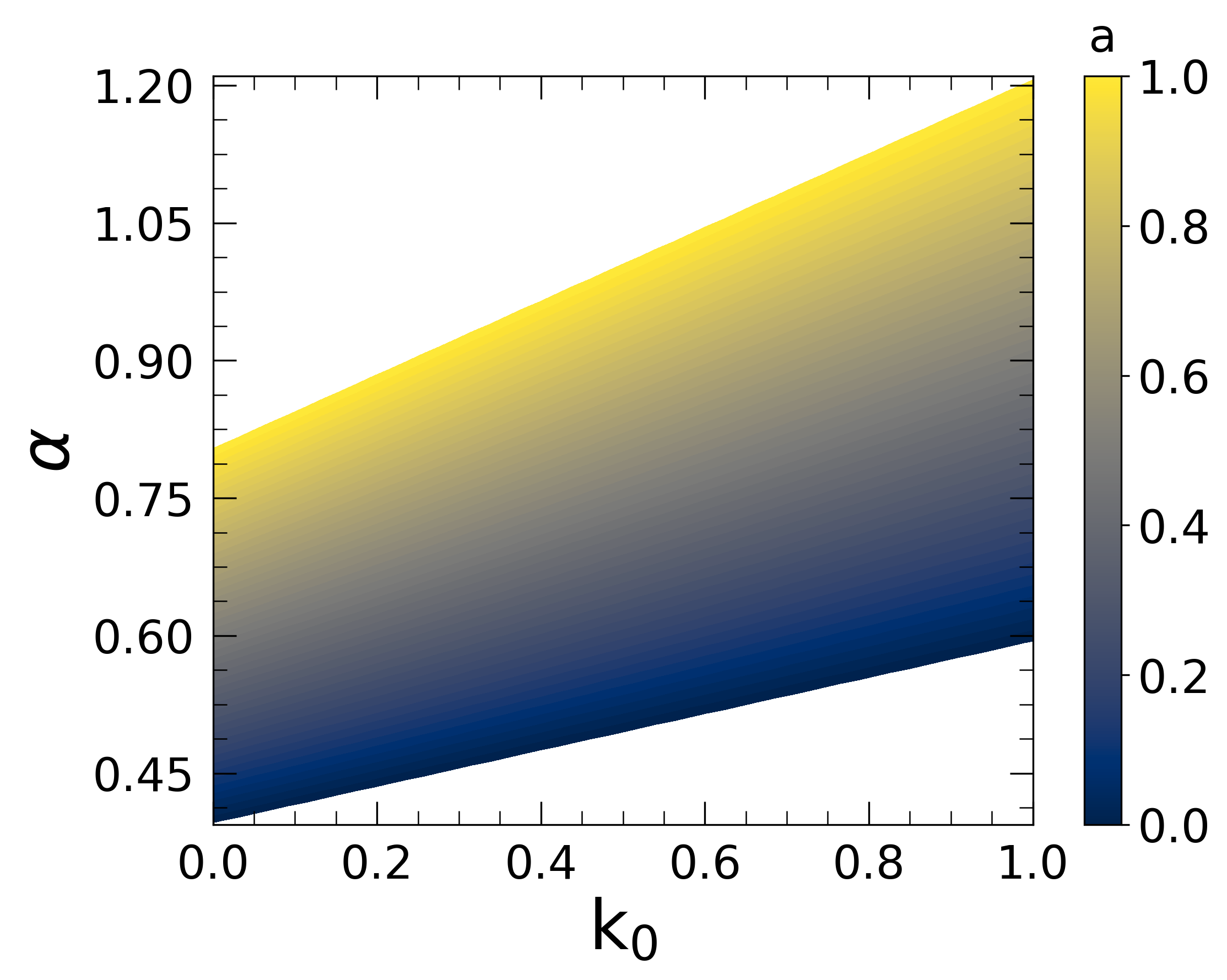}}
\subfigure(b){\includegraphics[width = 3.0in, height=2.7in]{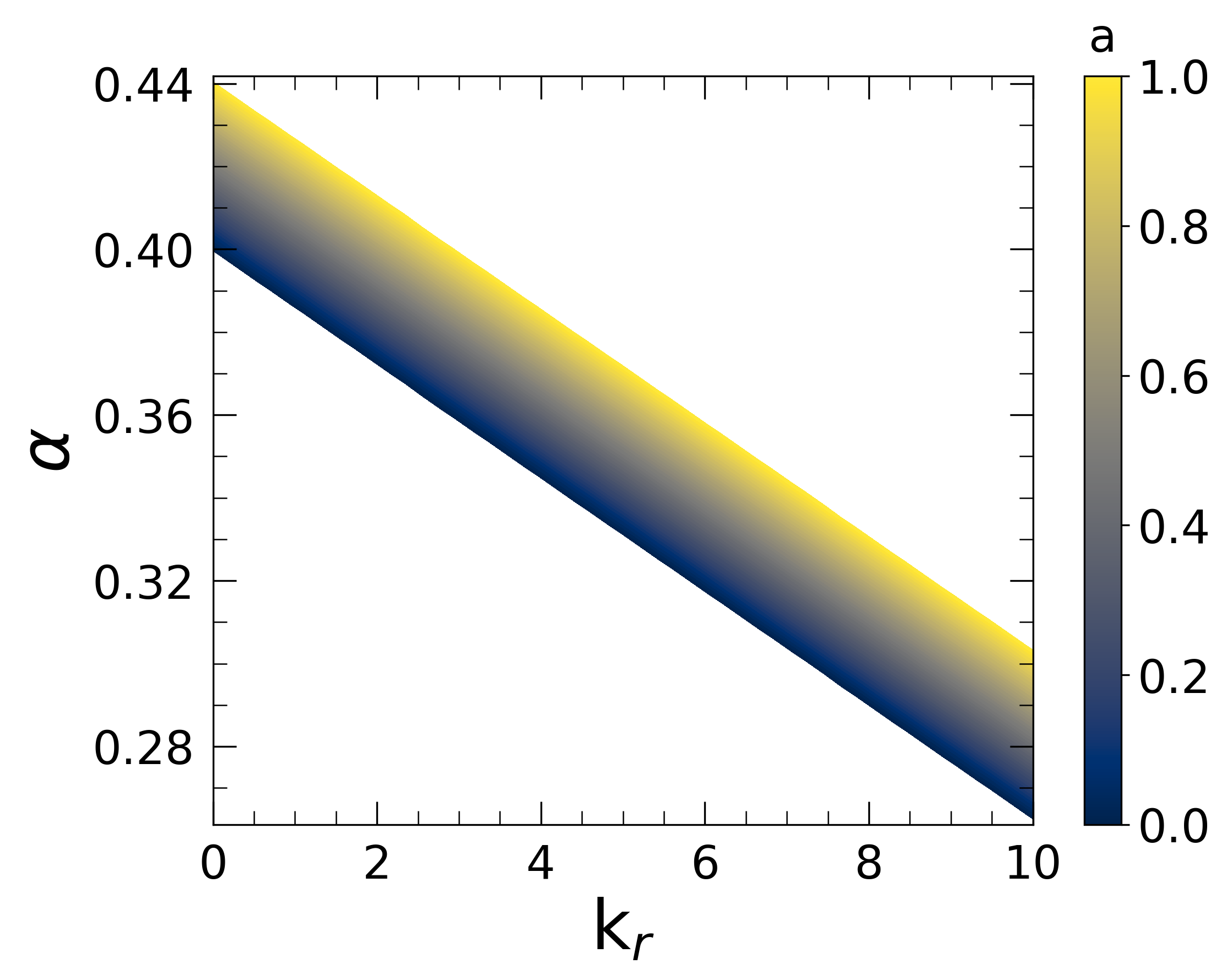}}
\caption{Weak deflection angle by Kerr black hole in plasma space-time as a function of plasma parameters ($k_0,k_r$) with the closest distance, $R=10$ for various spin parameters, with plasma profiles (a) $\omega_e^2=k_0\omega_o^2$, here the plasma parameter is $k_0$, and (b) $\omega_e^2=\frac{k_r}{r^2}\omega_o^2$, here the plasma parameter is $k_r$ and its value is taken accordingly to satisfy the low-plasma density and Eq. \ref{8}.}
\label{fig:5}
\end{figure*}
\begin{figure*}
\subfigure(a){\includegraphics[width = 2.9in, height=2.7in]{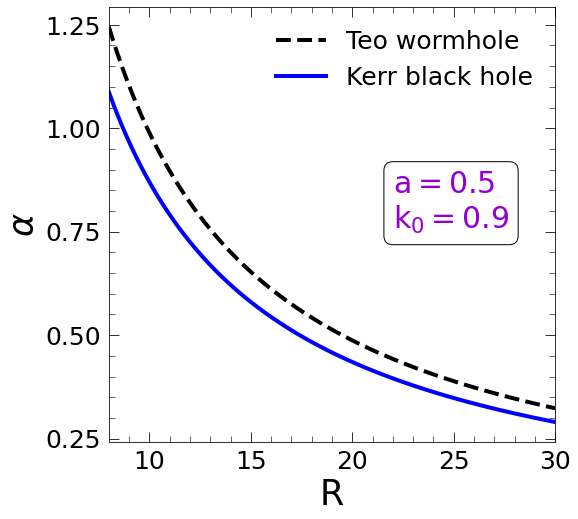}}
\subfigure(b){\includegraphics[width = 2.9in, height=2.7in]{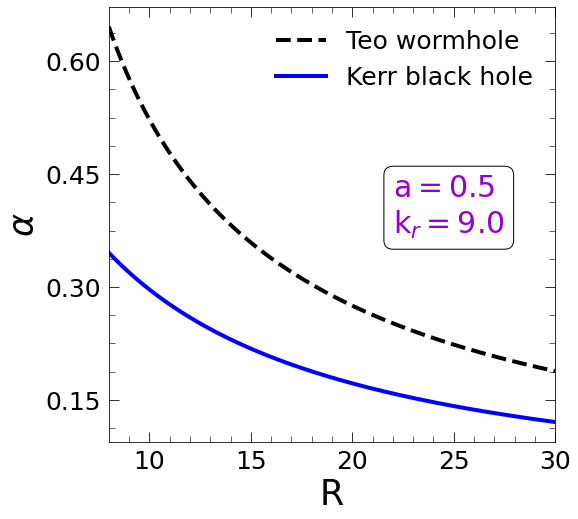}}
\caption{Weak deflection angle of Kerr black hole (solid blue curve) and Teo wormhole (dashed black curve) as a function of closest distance approach (R) for (a) uniform plasma distribution and (b) non-uniform plasma distribution.}
\label{fig:6}
\end{figure*}
\subsection{Uniform plasma distribution}
As discussed in the previous section, the uniform plasma density is taken as $ \omega_e^2=k_0\omega_o^2$. Now, let us expand expression \ref{44} with low plasma density in the context of a slow-rotating black hole within the weak field limit as,
\begin{equation}
    \bar{\alpha}  = 2\int_R^{\infty} \Phi_0 dr+2 \int_R^{\infty} \Phi_1 dr+2 \int_R^{\infty} \Phi_2 dr ,
\end{equation}
where,
\begin{align}
\begin{split}
\Phi_0 &= \frac{ R}{r \sqrt{r^2-R^2}}, \\
\Phi_1 &= \left(\frac{r^2(2a+R)+r R^2+R^3}{r^2 (r+R) \sqrt{r^2-R^2}}\right)\frac{m}{R}, \\
\Phi_2 &= \left(\frac{(a+R)}{(r+R) \sqrt{r^2-R^2}}\right)\frac{m k_0}{R}.
\end{split}
\end{align}
Solving the above integration will give the following expression for the weak deflection angle in the homogeneous plasma medium around the Kerr black hole,
\begin{equation}
    \alpha = \pm \left(\frac{4 m}{R}+ \frac{4 ma}{R^2} +\left(\frac{2m}{R}+\frac{2ma}{R^2}\right)k_0\right),
\end{equation}
where higher-order terms have been neglected. The values of $k_0$ considered in this study lie within the range of ($0,1$), as discussed in Section IV(A). Fig. \ref{fig:4}(a) illustrates the deflection angle as a function of the closest distance approach, conforming to the expected behavior of a decrease in deflection angle as we go farther away from the source since the effect of gravity becomes lesser at a farther distance, and here we are only considering the dispersive nature of plasma while neglecting its gravitational effect, so there are no anomalies from the general trend. Moreover, the deflection angle demonstrates an increasing trend with spin parameters and rising plasma density, as depicted in Fig. \ref{fig:5}(a) for the closest distance of $R=10$, and now we know why the shadow size increases as we increase the plasma density in the case of a homogeneous plasma profile, as shown in Fig. \ref{fig:2} (a). However, the result is somewhat surprising in the case of non-uniform plasma distribution, which will be discussed in the next subsection.

\subsection{Non-uniform plasma distribution}
The plasma profile for this case is considered as $\omega_e^2/\omega_o^2=k_r/r^2$ \cite{farruh2021}. Now, Let us expand Eq. \ref{44} for the non-homogeneous plasma scenario in the context of a slow-rotating black hole within weak field limit and low plasma density as,
\begin{equation}
    \bar{\alpha}  = 2\int_R^{\infty} \psi_0 dr+2 \int_R^{\infty} \psi_1 dr+2 \int_R^{\infty} \psi_2 dr,
\end{equation}
where,
\begin{align}
\begin{split}
\psi_0 &= \frac{R}{r\sqrt{r^2-R^2}}, \\
\psi_1 &= \Bigg(\frac{r^2+rR+ R^2}{ r^2 (r+R) \sqrt{r^2-R^2}}+ \\
&\quad +a\frac{2}{R (r+R) \sqrt{r^2-R^2}}\Bigg)m,\\
\psi_2 &= \Bigg(-\frac{1}{2rR \sqrt{r^2-R^2}}-\frac{2ma}{r^2R(r+R)\sqrt{r^2-R^2}}+\\
&\quad\frac{r^2+rR-R^2}{2r^2R^3 (r+R) \sqrt{r^2-R^2}}m\Bigg)k_r.
\end{split}
\end{align}
Now solving the above integration will give the following expression for the weak deflection angle in the presence of the non-uniform plasma distribution around the Kerr black hole,
\begin{equation}
    \alpha = \pm \left(\frac{4m}{R}+\frac{4ma}{R^2}-\left(\frac{2m}{R^3}+\frac{\pi}{2R^2}\right)k_r\right),
\end{equation}
here we have neglected the higher-order terms. Previous investigations have commonly explored plasma parameter values spanning from $0$ to $1$ \cite{Tsupko:2013cqa, farruh2021}. However, within the specified impact parameter constraints given by Eq. \ref{8} and the gravitational redshift condition given by Eq. \ref{6}, a range of plasma parameter values can be chosen to examine the deflection angle. It is essential to emphasize the significance of the redshift condition (Eq. \ref{6}) as it profoundly affects the path of light and consequently, the resulting deflection angle.

A noteworthy observation regarding radial plasma distributions is the decrease in the deflection angle with increasing plasma parameters as depicted in Figs. \ref{fig:4}(b) and \ref{fig:5}(b) for all spin parameter values. This observation presents a significant contrast to the behavior exhibited in the homogeneous case and provides a rationale for the observed negative impact of plasma on the shadow as illustrated in Fig. \ref{fig:2} (b). This finding contributes to our understanding of the interplay between plasma density, deflection angles, and the resulting shadow characteristics.

Finally, we have utilized the derived deflection angle results to compare our results of the Kerr black hole with the Teo wormhole \cite{kumar2023shadow}. In the case of homogeneous plasma, the deflection angles for both objects exhibit similarity when contrasted with the non-homogeneous plasma profile, as illustrated in Figs \ref{fig:6}(a) and \ref{fig:6}(b). This significant difference in deflection angles for the non-uniform plasma scenario adds to the intrigue of studying plasma distribution to distinguish between these two compact objects. Such investigations serve as valuable tools for comprehending the plasma properties surrounding these objects and enabling a clear differentiation between the wormhole and the black hole.

\begin{figure*}[h]
\subfigure{\includegraphics[width = 3.3in, height=2.8in]{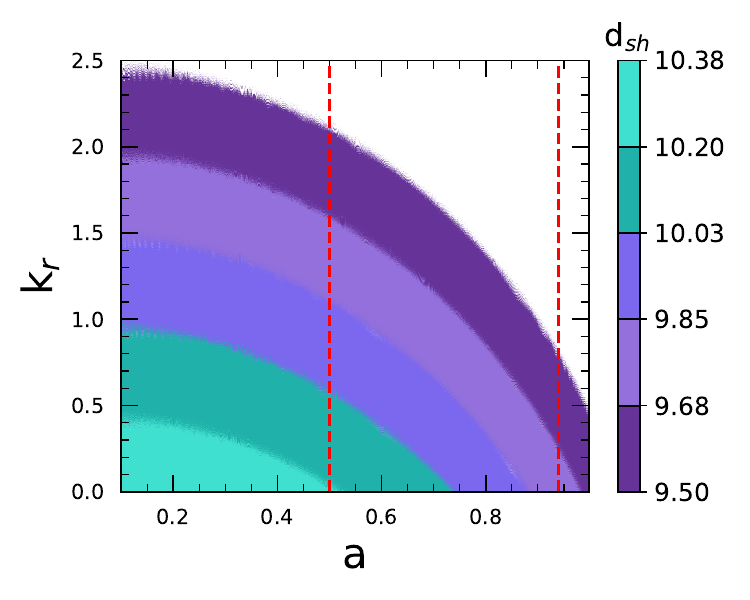}}
\subfigure{\includegraphics[width = 3.3in, height=2.8in]{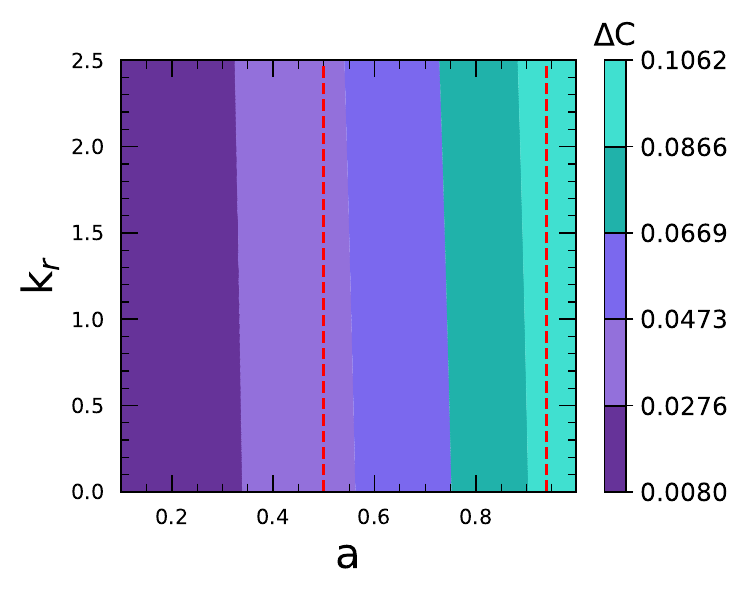}}
\caption{ Dependence of the angular size and the deviation of the shadow on the radial plasma parameter ($k_r$) and spin with $\theta=17^\circ$. The two red dashed lines indicate the spin range of $0.5\leq a \leq 0.94$. Considering plasma profile: $\omega_e^2=\frac{k_r\sqrt{r}}{r^2+a^2\cos^2\theta}\omega_o^2$.}
\label{fig:7}
\end{figure*}

\begin{figure*}[h]
\subfigure{\includegraphics[width = 3.3in, height=2.8in]{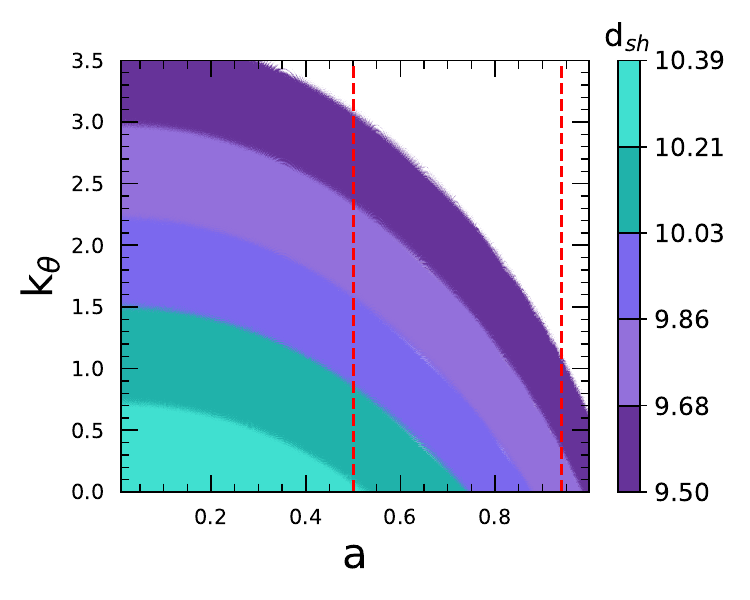}}
\subfigure{\includegraphics[width = 3.3in, height=2.8in]{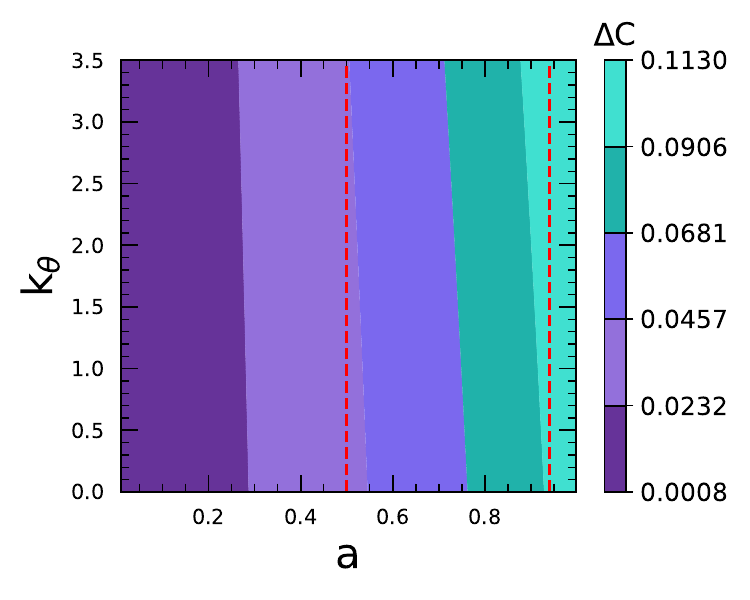}}
\caption{Dependence of the angular size and the deviation of the shadow on longitudinal plasma parameter ($k_\theta$) and spin with $\theta=17^\circ$. The two red dashed lines indicate the spin range $0.5 \leq a \leq 0.94$. Considering plasma profile: $\omega_e^2=k_\theta\frac{1+2\sin^2\theta}{r^2+a^2\cos^2\theta}\omega_o^2$ .}
\label{fig:8}
\end{figure*}

\section{CONSTRAINING THE PLASMA PARAMETERS}\label{Sec6}

In this section, we will constrain the plasma parameter with the help of previous analyses \cite{Rahaman_2021, tamburini2019measurement, Nemmen_2019} and the observed shadow data provided by the EHT for the supermassive black hole M87* \cite{EHT2019_1}. Specifically, we utilize the average angular size of the shadow and its deviation from circularity as constraints. As the shadow exhibits reflection symmetry with respect to the $\alpha$-axis, we can find its geometric center $\left(\alpha_0, \beta_0\right)$ by using the expressions $\alpha_0=1 / A \int \alpha dA$ and $\beta_0=0$. Here, $dA$ represents an area element. Additionally, we introduce an angle $\phi$ between the $\alpha$-axis and the vector connecting the geometric center with a point $(\alpha, \beta)$ on the shadow's boundary in the celestial plane. Consequently, the average radius $R_{av}$ of the shadow can be determined using the following expression \cite{Rahaman_2021},
$$
R_{a v}^2=\frac{1}{2 \pi} \int_0^{2 \pi} l^2(\phi) d \phi,
$$
where $l(\phi)=\sqrt{\left(\alpha(\phi)-\alpha_c\right)^2+\beta(\phi)^2}$ and $\phi=\tan ^{-1}\left(\beta(\phi) / \left(\alpha(\phi)-\alpha_c\right)\right)$. Following \cite{Rahaman_2021}, we define the deviation $\Delta C$ from circularity as,
$$
\Delta C=\frac{1}{R_{a v}} \sqrt{\frac{1}{2 \pi} \int_0^{2 \pi}\left(l(\phi)-R_{a v}\right)^2 d \phi},
$$
Here $\Delta C$ denotes the fractional root mean square (RMS) distance from the average radius of the observed shadow. As reported by the EHT collaboration \cite{EHT2019_1}, the angular size of the observed shadow is measured to be $\Delta \theta_{\text{sh}} = 42 \pm 3 \mu$as with the deviation $\Delta C$ found to be less than $10\%$. Additionally, following the findings of \cite{EHT2019_1}, we adopt a distance to M87* of $D = (16.8 \pm 0.8)$ Mpc and a mass for the object of $M = (6.5 \pm 0.7) \times 10^9 M_{\odot}$. Based on these measurements, we can deduce the average diameter of the shadow can be given by,
$$
\frac{d_{s h}}{M}=\frac{D \Delta \theta_{s h}}{G M}=11.0 \pm 1.5.
$$
\begin{table*}[ht]
\addtolength{\tabcolsep}{25pt}
\fontsize{13.5pt}{10.25pt}
\centering
\renewcommand{\arraystretch}{2.5}
\caption{Constrained value of plasma parameters ($k_r, k_\theta$) using the shadow of M87* black hole as discussed in Sec. \ref{Sec6}}
\begin{tabular}{|c|c|}
\hline
\text{\fontsize{12}{14}\selectfont Plasma distributions} & \text{\fontsize{12}{14}\selectfont Plasma parameters ($k_r, k_\theta$)} \\
\hline
$\omega_e^2=\frac{k_r\sqrt{r}}{r^2+a^2\cos^2\theta}\omega_o^2$ &  \text{\fontsize{12}{12}$0.75 \leq k_r \leq 2.05$} \\

 $\omega_e^2=k_\theta\frac{1+2\sin^2\theta}{r^2+a^2\cos^2\theta}\omega_o^2$ & \text{\fontsize{12}{12}$ 1.05 \leq k_\theta \leq 3.05$} \\
\hline
\end{tabular}

\label{Table 1}
\end{table*}

Considering the errors added in quadrature, the quantity $\frac{2R_{\text{av}}}{M}$ represents the average diameter of the shadow. Figs. \ref{fig:7}(a) and \ref{fig:8}(a) depict the average diameter of the shadow, while Figs. \ref{fig:7}(b) and \ref{fig:8}(b) illustrate the deviation from circularity for various values of the plasma parameters $k_r$ and $k_\theta$ as well as the spin parameter. In our analysis, we have chosen an inclination angle of $\theta_o = 17^\circ$ as the inclination angle from the jet axis of the M87* \cite{EHT2019_1}. We have also utilized the conclusions provided in \cite{tamburini2019measurement, Nemmen_2019} for the spin parameter to further constraint the plasma parameters, where the spin parameter falls within the range $0.5 \leq a \leq 0.94$, which we have shown by the vertical red lines in Figs. \ref{fig:7} and \ref{fig:8}. We have constrained the plasma parameters using the lower bound of the shadow size at $9.5$, and by considering the spin constraints, we find that the value of the plasma parameter lies within $0.75 \leq k_r \leq 2.05$ (see Fig. \ref{fig:7}(a)). Furthermore, Fig. \ref{fig:7}(b) suggests that the upper limit for the deviation from circularity is below $10\%$.

We have also tried to constraint the latitudinal plasma parameter ($k_\theta$) using the lower bound of shadow size of $9.5$ and using the spin constraints; the value of the plasma parameter lies within $1.05 \leq k_\theta \leq 3.05$ (see Fig. \ref{fig:8}(a)), while it can be concluded from Fig. \ref{fig:8}(b) that the upper limit of circularity deviation is less than $10\%$.
These constrained plasma parameters ($k_r, k_\theta$) have been summarized in Table \ref{Table 1}.
Please note that we have tried to constrain the plasma parameters, which are discussed in Section IV.

\section{Conclusion}
The primary objective of this investigation was to explore the characteristics of the shadows cast by the Kerr black hole in the presence of plasma and compare them analytically with those of the wormhole in a uniform plasma space-time. Additionally, we aimed to examine the deflection angle of light in the plasma medium surrounding the Kerr black hole and compare it with that of the rotating wormhole. Ultimately, our goal was to apply these findings to constrain the plasma parameters using M87* black hole results. By analyzing the interplay between plasma and the gravitational field, we sought to deepen our understanding of these intriguing astrophysical phenomena and shed light on the nature of compact objects in the universe.


By considering specific plasma density conditions outlined in Section III, we derived the boundary expressions for shadows in the plasma case, where $\Omega_r=0$ and $\Omega_\theta=0$, enabling a comparison with the vacuum case. Our findings revealed a negative impact on the shadow in the non-uniform plasma scenario, while the uniform plasma exhibited a positive impact, with the shadow disappearing for higher plasma densities. Furthermore, we proceeded to compare the shadows of the black hole and wormhole in the homogeneous plasma, which exhibited similarities to the vacuum case but with larger shadow sizes. As a result, the comparison of their shadows in the homogeneous plasma context remained consistent with the vacuum case.
Through a meticulous investigation of light propagation in the presence of plasma, we aimed to distinguish between the Kerr black hole and the rotating wormhole. In Section V, we performed comprehensive calculations to determine the deflection angle by the Kerr black hole in plasma space-time. The gravitational lensing phenomenon bears significant implications for astrophysical observations, and our study sheds new light on the influence of plasma on the deflection angle. Notably, we made intriguing observations: as the plasma parameter increases, the deflection angle decreases in non-uniform plasma media, contrary to the behavior observed in homogeneous plasma profiles. Moreover, we conducted a comparative analysis of the deflection angles between the black hole and the wormhole. Intriguingly, we found that in the case of homogeneous plasma, these compact objects appear indistinguishable. However, in the non-uniform plasma case, a significant disparity in the deflection angles emerges. This captivating outcome underscores the importance of further investigation into the observational aspects and exploration of the plasma distribution surrounding compact objects, aiding in their differentiation.

Lastly, we applied our analyses to constrain the plasma parameters. Our analysis yielded the radial plasma parameter, $k_r$, in the range of 0.75 to 2.05, and the latitudinal plasma parameter, $k_\theta$, in the range of 1.05 to 3.05. These ranges were determined by imposing constraints on the shadow diameter, ensuring it fell within the minimum allowed size, as well as considering the permissible range of spin parameters.

The future low-frequency observations can use our comparison between a wormhole and a black hole to distinguish them from each other however the real environment around such compact objects highly deviates from the current scope of this paper, nevertheless, such studies can help to understand the qualitative nature of the different compact objects. The boundary of the shadow depends on the space-time (considering the geodesics are travelling in the plasma medium in our case) and the gravitational strength of the object \cite{Synge:1966okc, Luminet:1979nyg, Shaikh:2018kfv}. Therefore, in such a scenario, we can understand the qualitative difference between the BH and WH.
In our future research endeavors, we plan to delve deeper into the effects of non-uniform plasma densities on both the Kerr black hole and the Teo wormhole. We also plan to compare the results of the parametrized metrics for the black hole by Konoplya et al. \cite{Konoplya:2016jvv} with the parameterized wormhole metric by Bronnikov et al.\cite{Bronnikov:2021liv} within the plasma space-time. By examining their behavior under the same plasma profile, we aim to enhance our understanding of their characteristics in more realistic conditions. This investigation will contribute to our comprehension of the intricate interplay between plasma and the propagation of light rays, offering valuable insights into the behavior of these fascinating astrophysical objects.



\section*{Acknowledgement}
The work of SC is supported by Mathematical Research Impact Centric Support (MATRICS) from the Science and Engineering Research Board (SERB) of India
through grant MTR/2022/000318. We thank the anonymous referees for various useful comments, which helped us to raise the quality of the present manuscript.

\bibliography{references}
\end{document}